\documentstyle[aps]{revtex}
\draft

\begin{document}
\author{Wei-Min Sun, Xiang-Song Chen and Fan Wang}
\address{Department of Physics and Center for Theoretical Physics, Nanjing 
University, Nanjing 210093, China}
\title{A Simple Proof That $C^{\infty}({\bf R}^n,U(1))$ Does Not Have a Haar Measure}
\maketitle

\begin{abstract}
We give a simple proof that there does not exist a Haar measure on the group $C^{\infty}({\bf R}^n,U(1))$.
\end{abstract}

Nowadays functional integral has become an indispensible tool in quantum field theory.
The most common type of functional integral encountered is functional integral in field space. The rigorous mathematical formulation of it requires the construction
of a measure in some function space. In the framework of Euclidean field theory 
this has been done for the case of free fields and some interacting fields 
in low spacetime dimensions. \cite{Glimm}. The measures constructed are of the form
$e^{-S[\phi]}\prod_x d\phi(x)$ where $S[\phi]$ represents the Euclidean action.
The presence of the factor $e^{-S[\phi]}$ is essential for defining a true measure in field space(the formal expression $\prod_x d\phi(x)$ is not a true measure; in
fact it is known that there does not generally exist a translationally invariant measure in an infinite dimensioinal vector space).

Apart from functional integral on field space, in various applications we have another type of functional integral: functional integral over the local gauge group. The most famous example is the Faddeev-Popov
method \cite{Faddeev} of quantizing a gauge field theory. In the F-P method and some other contexts \cite{Polonyi} it is assumed that there exists a Haar measure of the form $\prod_x d\omega(x)$ on the local gauge group, where $d\omega(x)$ stands for the usual Haar measure on some compact Lie group(at the spacetime point $x$). But 
we know that the local gauge group is infinite dimensional and not locally compact and hence the existence of a translationally invariant measure is not guaranteed. In this paper we give a simple proof that on the group $C^{\infty}({\bf R}^n,U(1))$ there does not exist a translationally invariant bounded positive measure.
 
Let $G$ denote the group $C^{\infty}({\bf R}^n,U(1))$ endowed with the $C^{\infty}$-topology \cite{Milnor}. Let $\mu$ be a bounded positive measure on the Borel $\sigma$-field of $G$. Let $f \in C^{\infty}({\bf R})$ and satisfy the following two conditions:
\begin{equation}
|f(u)| \leq K, \forall u \in {\bf R}
\end{equation}
\begin{equation}
0< f'(u) \leq K', \forall u \in {\bf R}
\end{equation}
where $K$ and $K'$ are two positive constants.
One such example is $f(u)=\int_{-\infty}^{u} dt e^{-a^2 t^2}, a \in {\bf R}-\{0\}$.
Let $F(\omega)$ be the following real-valued function on $G$:
\begin{equation}
F(\omega)=f(-i\omega^{-1}(x)\partial_\mu\omega(x)), \forall \omega \in G
\end{equation}
where $x$ is a fixed point in ${\bf R}^n$.
Obviously $F(\omega)$ is a continous and bounded function on $G$, and hence is $\mu$-integrable. Suppose $\mu$ is translationally invariant. Then we have:
\begin{equation}
\int_G D\mu(\omega) F(\omega)=\int_G D\mu(\omega) F(\omega_0\omega), \forall \omega_0 \in G
\end{equation}
Using (3) we have
\begin{equation}
\int_G D\mu(\omega)f(-i \omega^{-1}(x)\partial_\mu\omega(x))= \int_G D\mu(\omega)
f(-i\omega^{-1}(x)\partial_\mu\omega(x)-i\omega^{-1}_0(x)\partial_\mu\omega_0(x)),
\forall \omega_0 \in G
\end{equation}
From the arbitariness of $\omega_0$ and the surjectiveness of the mapping $\omega
\mapsto -i\omega^{-1}(x)\partial_\mu\omega(x)$, we obtain:
\begin{equation}
\int_G D\mu(\omega)f(-i\omega^{-1}(x)\partial_\mu\omega(x))=\int_G D\mu(\omega)
f(-i\omega^{-1}(x)\partial_\mu\omega(x)+c), \forall c \in {\bf R}
\end{equation}
The RHS of (6) is independent of $c$. Differentiating with respect to $c$ at $c=0$ gives
\begin{equation}
\frac{d}{dc}|_{c=0}\int_G D\mu(\omega) f(-i\omega^{-1}(x)\partial_\mu\omega(x)+c)=0
\end{equation}
Now consider the following family of real-valued continuous functions on $G$:
\begin{equation}
F(\omega;\eta)=\frac{f(-i\omega^{-1}(x)\partial_\mu\omega(x)+\eta)-f(-i\omega^{-1}(x)\partial_\mu\omega(x))}{\eta}, \eta \in {\bf R}
\end{equation}
From (2) and the Lagrange mean-value theorem we see that
\begin{equation}
|F(\omega;\eta)| \leq K', \forall \omega \in G, \eta \in {\bf R}
\end{equation}
The family $\{F(\omega;\eta);\eta \in {\bf R}\}$ is $\mu$-integrable with respect to
$\omega$ and converges everywhere to $f'(-i\omega^{-1}(x)\partial_\mu\omega(x))$ when $\eta \rightarrow 0$:
\begin{equation}
\lim_{\eta \rightarrow 0}F(\omega;\eta)=f'(-i\omega^{-1}(x)\partial_\mu\omega(x)),\forall
\omega \in G
\end{equation}
From the dominated convergence theorem, we have
\begin{eqnarray}
\lim_{\eta \rightarrow 0}\int_G D\mu(\omega) F(\omega;\eta)&=& \int_G D\mu(\omega)\lim_{\eta \rightarrow 0}F(\omega;\eta) \nonumber \\
&=& \int_G D\mu(\omega) f'(-i\omega^{-1}(x)\partial_\mu\omega(x))
\end{eqnarray}
Comparing with (7) and (8) we get
\begin{equation}
\int_G D\mu(\omega) f'(-i\omega^{-1}(x)\partial_\mu\omega(x))=0
\end{equation}
But this equilty is impossible because the integrand is greater than 0 evereywhere. 
So we conclude that on the group $C^{\infty}({\bf R}^n,U(1))$(endowed with the $C^{\infty}$-topology) there does not exist a translationally invariant bounded positive measure defined on the Borel $\sigma$-field of $G$. Especially the formal
expression $\prod_x d\omega(x)$(with the normalization $\int_{U(1)}d\omega(x)=1$) 
does not define a true Haar measure on the local gauge group $C^{\infty}({\bf R}^n,
U(1))$.

We thank Feng-Wen An for reading the manuscript. This work is supported by NSF, SED and SSTC of China.

\end{document}